\documentclass[amsmath,amssymb,floatfix,twocolumn,showpacs]{revtex4}
    \usepackage{graphicx}       \usepackage{bm}
    \usepackage{epsf}           \usepackage{epsfig}
\newcommand{\bra}[1]{\mbox{$\left\langle #1 \right|$}}
\newcommand{\ket}[1]{\mbox{$\left|#1\right\rangle$}}

\newcommand{\be}{\begin{equation}}  \newcommand{\ee}{\end{equation}}
\newcommand{\bq}{\begin{eqnarray}}  \newcommand{\eq}{\end{eqnarray}}
       \newcommand{\tr}{\text{tr}}
          
\newcommand{\up}{\uparrow}          \newcommand{\down}{\downarrow}

\begin{document}
\title{Chiral entanglement in triangular lattice models}

\author{Dimitris I. Tsomokos$^1$, Juan Jos\'e Garc{\'\i}a-Ripoll$^2$, Nigel
R. Cooper$^3$ and Jiannis K. Pachos$^4$}
\affiliation{$^1$ Quantum Physics Group, STRI, School of Physics,
Astronomy \& Mathematics, University of Hertfordshire, Hatfield AL10 9AB,
UK
\\$^2$Departamento de F{\'\i}sica Te\'orica I, Universidad Complutense,
Madrid, 28040, Spain
\\$^3$Theory of Condensed Matter Group, Cavendish Laboratory, J.J.
Thomson Avenue, Cambridge CB3 0HE, UK
\\$^4$School of Physics \& Astronomy, University of
Leeds, Leeds LS2 9JT, UK}
\date{\today}

\begin{abstract}
We consider the low energy spectrum of spin-$\frac{1}{2}$ two-dimensional
triangular lattice models subject to a ferromagnetic Heisenberg interaction
and a three spin chiral interaction of variable strength. Initially, we
consider quasi-one dimensional ladder systems of various geometries.
Analytical results are derived that yield the behavior of the ground states,
their energies and the transition points. The entanglement properties of the
ground state of these models are examined and we find that the entanglement
depends on the lattice geometry due to frustration effects. To this end, the
chirality of a given quantum state is used as a witness of tripartite
entanglement. Finally, the two dimensional model is investigated numerically
by means of exact diagonalization and indications are presented that the low
energy sector is a chiral spin liquid.
\end{abstract}
%----------------------------------------------------------
 \maketitle

%%%%%%%%%%%%%%%%%%%%%%%%%%%----------- new sec ------------
\section{Introduction}

Exotic quantum orders of correlated many-body systems have been
studied extensively over the last years \cite{Wen_book}. Prime
examples include the Laughlin states \cite{Laughlin_1983}, which
were studied in relation to the fractional quantum Hall effect,
and the chiral spin states \cite{Wilczek_1989} that were studied
in relation to high-$T_c$ superconductivity. Chiral spin states
break spontaneously the parity (P) and time-reversal (T)
invariance. It is known that in two spatial dimensions it is
possible for such ``incompressible'' quantum systems with a mass
gap to exhibit \emph{topological order} \cite{Wen_Niu}. One of the
most intriguing properties in this case is that the degree of the
ground state degeneracy depends on the topology of the surface in
which the system resides. As a result, topologically ordered
quantum states cannot be distinguished by any local measurements
but are only globally distinct. The outstanding problem of
characterizing quantum states that possess topological order
\cite{Wen_2002} has recently motivated the study of the
many-particle entanglement properties of such states
\cite{S_topo}.

In this paper we consider a model that breaks the PT-symmetry
\emph{explicitly} via a scalar chiral interaction term
\begin{equation}\label{chirality_operator}
{\cal X}= \vec{\sigma}_i \cdot \vec{\sigma}_j \times
\vec{\sigma}_k
\end{equation}
while it preserves the SU(2) rotational symmetry. We study the
quantum phase transitions of this model and the properties of its
different phases. In particular, we are interested in the local
indistinguishability of the degenerate ground states, and the
absence of any structure in the classical correlations
\cite{Wen_2002,Lhuillier}. In addition, we introduce an
entanglement witness that quantifies the quantum correlations of
the system due to the chiral currents. This is a suitable
multipartite entanglement witness that detects two-body and
different classes of three-body entanglement.

From the experimental point of view, measurements of the scalar chirality in
pyrochlore ferromagnets \cite{experiment_1} and Kagom{\'e} lattice
structures \cite{experiment_2} have been performed. Moreover, recent
theoretical and experimental advances in cold atomic and molecular physics
indicate the possibility of generating the required chiral interactions. As
we shall see, the chiral model studied here could be implemented with cold
atoms superposed by optical lattices in the presence of effective magnetic
fields \cite{Pachos,Kay,Pachos_Rico,Jaksch,Zoller}.

The model we study is a two dimensional triangular lattice of
spin-$\frac{1}{2}$ particles subject to chiral and ferromagnetic
Heisenberg interactions. The Hamiltonian of the system is given by
\begin{equation}\label{Ham}
{\cal H}(\lambda) = -\sum_{\langle i,j \rangle}
\vec{\sigma}_{i}\cdot \vec{\sigma}_j + \lambda \sum_{\langle i,j,k
\rangle} \vec{\sigma}_i \cdot\vec{\sigma}_j \times\vec{\sigma}_k
\end{equation}
where $\langle i,j \rangle$ and $\langle i,j,k \rangle$ denote any two
and three nearest neighboring sites, respectively, and $\vec{\sigma}_i
\equiv (X_i,Y_i,Z_i)$ where $X_i,Y_i,Z_i$ are the Pauli matrices for
spin $i$. The parameter $\lambda$ is a real number that determines the
relative strength of the chiral and ferromagnetic interactions. The
sign of $\lambda$ is irrelevant for the eigenvalues and it can be
changed with a time reversal transformation.

The ferromagnetic and chiral terms favor different orders,
although they are both $SU(2)$ symmetric. When $|\lambda| \ll 1$,
the Hamiltonian (\ref{Ham}) has a ferromagnetic ground state which
is a product state with all spins pointing along the same
direction. On the contrary, when $|\lambda| \gg 1$, the chiral
part favors a ground state which is entangled, typically belongs
to the singlet sector and has zero magnetization and nonzero
chirality.

The focus of our work is on the properties of the chiral phase and
on the phase transition that takes place as the value of
$|\lambda|$ is changed. Our study begins in
Sect.~\ref{sec:entanglement} with an interpretation of the
chirality for three spin systems, focusing on the relation between
chirality and entanglement. In Sect.~\ref{sec:ladder} we introduce
simple quasi-1D geometries, such as regular polygons and spin
ladders, and analyze the ground state both numerically and
analytically. We present indications of a topological quantum
phase transition and study how different levels of frustration on
the chirality give rise to ground states with different kinds of
entanglement. In Sect.~\ref{sec:torus} we move to a more
sophisticated geometry, namely a 2D lattice with periodic boundary
conditions which forms a torus. Here we perform a numerical study
of the phase transition and characterize what we conjecture to be
a spin-liquid order in the chiral phase. Finally, as commented
above, Sect.~\ref{sec:implementation} discusses the possibility of
realizing this model using cold atoms in optical lattices, and we
summarize our conclusions in Sect.~\ref{sec:summary}.

%%%%%%%%%%%%%%%%%%%%%%%%%%%----------- new sec ------------
\section{Chirality and entanglement}

\label{sec:entanglement}

Consider a system of three spins in a triangular configuration. We
pose the following question: is there any relation between
entanglement properties of a pure state of these spins,
$\ket{\Psi}$, and their scalar chirality,
\begin{equation}
\chi =\bra{\Psi} {\cal X} \ket{\Psi}, \label{chirality}
\end{equation}
defined as the expected value of the operator in
Eq.~(\ref{chirality_operator})? In subsections B and C below we answer this
question and show that indeed the chirality can be used to detect the
existence of two- and three-partite entanglement in arbitrary states. This
is true for any number of spins, which means that by computing the chirality
on different sets of spins one can learn about the distribution of
entanglement in the ground state of a Hamiltonian, such as the one of Eq.
(\ref{Ham}) (\emph{cf.} Sect.~\ref{sec:ladder}). Clearly, the scalar
chirality can also be defined for general mixed states, $\rho$, as $\tr
(\rho {\cal X})$. This definition is employed in Sections \ref{sec:ladder}
and \ref{sec:torus} to characterize the multiqubit ground states of quasi-1D
or 2D systems.

\subsection{Eigenstates and eigenvalues}

\begin{table}
  \centering
  \begin{tabular}{l|l|l}
    $\ket{S,S_z}$ & State & $\chi$ \\
    \hline \hline
    $\ket{\tfrac{3}{2},\tfrac{3}{2}}$ &
    $\ket{\uparrow\uparrow\uparrow}$
    & 0 \\
    $\ket{\tfrac{3}{2},\tfrac{1}{2}}$ &
    $\tfrac{1}{\sqrt{3}}(\sigma^x_1 + \sigma^x_2 +
    \sigma^x_3) \ket{\uparrow\uparrow\uparrow}$ &
    $0$ \\
    $\ket{\tfrac{1}{2},\tfrac{1}{2}}^{+}$ &
    $\tfrac{1}{\sqrt{3}}(\sigma^x_1 + \omega \sigma^x_2 +
    \omega^2 \sigma^x_3) \ket{\uparrow\uparrow\uparrow}$ &
    $-2\sqrt{3}$ \\
    $\ket{\tfrac{1}{2},\tfrac{1}{2}}^{-}$ &
    $\tfrac{1}{\sqrt{3}}(\sigma^x_1 + \omega^* \sigma^x_2 +
    \omega^{*2} \sigma^x_3) \ket{\uparrow\uparrow\uparrow}$ &
    $+2\sqrt{3}$
  \end{tabular}
  \caption{
    Half of the eigenstates of the chirality operator of Eq.
    (\ref{chirality_operator}) for three spins expressed
    in terms of the phase $\omega=\exp(2\pi i/3)$
    and its complex conjugate $\omega^{*}$.
    The eigenstates with negative values of $S_z$ are obtained by
    flipping all spins.
  }
  \label{tab:eigenstates}
\end{table}

The chirality operator ${\cal X}$ is Hermitian, imaginary
---i.~e. changes sign under complex conjugation--- and it is
invariant under global rotations. The Hermitian nature implies that $\chi$
is real, while the rotational symmetry implies that we can split its
eigenstates into the eigenstates of $S^2$ and $S_z$.

From the relation ${\cal X}^2 = -(\vec{\sigma}_1 + \vec{\sigma}_2
+\vec{\sigma}_3)^2+15$ we deduce that the chirality operator can only take
values between $+2\sqrt{3}$ and $-2\sqrt{3}$. This relation is further
exploited in Ref.~\cite{Wilczek_1989} to compute all eigenstates and
eigenvalues, which we summarize in Table~\ref{tab:eigenstates}. The
eigenspace of ${\cal X}$ decomposes into a spin-$\tfrac{3}{2}$ multiplet
with eigenvalue $\chi=0$, and two different spin-$\tfrac{1}{2}$ with
eigenvalues $\chi=\pm 2\sqrt{3}$, so that in total we have $4+2+2=8$
eigenstates, as expected.

Remarkably, the states with nonzero chirality are obtained when a
single spin is flipped with a relative phase $\omega$ or
$\omega^*$ between sites. Such a wavefunction can be interpreted
as a current moving left- or rightwards thus honoring the name
``chirality'' of the operator. This point is even more evident in
the bosonic model developed in Sect.~\ref{fermionization}, but it
is also a signature of the relation between chirality and W-like
entangled states \cite{Duer}.

\subsection{Three-partite states and chirality}
\label{entanglement}

The space of states for three spins (or qubits) is a particular case in
which all states can be classified into few classes depending on their
entanglement properties \cite{Duer}. In this section we construct explicitly
the most relevant instances of each class and analyze the corresponding
values of the chirality. The goal is to find a relation between chirality
and pure three-partite entanglement.

To begin with, we consider product states of the form $\ket{\Psi}
=\ket{\up}\otimes (a \ket{\up} + b\ket{\down}) \otimes (a'\ket{\up}
+b'\ket{\down})$, where without loss of generality the first spin is
oriented upwards. We find that $\chi=-4 {\rm Im} (a^*ba'b'^*)$ becomes
maximum when $|a|=|b|=|a'|=|b'|=1/\sqrt{2}$ and the phases are suitably
chosen. Thus, for a product state we will always have $|\chi|\le 1.$ If the
chirality takes a value larger than $1$, it signals the presence of quantum
correlations between two or more spins.

We now consider bipartite entangled states of the form
$\ket{\Psi_{12}} \otimes \ket{\Psi_3}$. Due to rotational symmetry
we may assume $\ket{\Psi_3}=\ket{\up}$ and $\ket{\Psi_{12}} =
a\ket{\up\up} + b\ket{\up\down} + c\ket{\down\up} +d
\ket{\down\down}$. Consequently, we obtain $\chi = 2i(bc^* -
b^*c)$, with a maximum absolute value given by $|\chi| = 4|bc|
\leq 2$. It is noted that, in this case, $|\chi|$ is equal to
twice the concurrence of the $\ket{\Psi_{12}}$ state~\cite{Hill}. To see
that this is not accidental consider the state $\ket{\Psi_{12}} =
\alpha\ket{\up\up} + \beta\ket{\down\down}$. The local unitaries that
optimize the expectation value of the relevant part of the chiral operator,
$X_1Y_2-Y_1X_2$, implement the rotations $X_2\rightarrow -Y_2$ and $Y_2
\rightarrow X_2$. Thus, the maximum expectation value of the chiral operator
corresponds to the expectation value of $2Y_1Y_2$, which is twice the
concurrence. Continuing with our reasoning, we have found that the maximum
chirality bipartite entangled states can give is $\chi = 2$. If the scalar
chirality assumes a larger value, then genuine three-party entanglement must
be present.

We finally arrive to the three-partite entangled states, which can either be
GHZ-like or W-like states \cite{Duer}. In the case of the GHZ-like state
$\ket{\Psi} = a\ket{\up\up\up} +b\ket{\down\down\down}$, we obtain $|\chi|
=|ab| 3\sqrt{3}$ with maximal value $\chi=3\sqrt{3}/2>2$. In the case of
W-like states the maximum value of $\chi$ is attained for the eigenstate of
the chirality operator with the maximal eigenvalue,
$\ket{\frac{1}{2},\frac{1}{2}}^+$, which gives $\chi = 2\sqrt{3}$ (see
Table~\ref{tab:eigenstates}). Thus, we can employ the chirality of three
spins as an observable to distinguish between the different types of three
partite entanglement.

It is noted that for states which are symmetric under the exchange of any
two spins, 1, 2 or 3, the above criteria simplify. Namely, it can be proven
that the chirality $\chi$ is different from zero only if the state possesses
genuine tripartite entanglement.

\subsection{Chirality operator as an entanglement witness}

We showed in the previous subsection that for all separable states
$\ket{\Psi}$ we have $|\langle \Psi |{\cal X}| \Psi\rangle | = |\chi| \le
1$. States whose chirality is greater than $1$ necessarily have some amount
of entanglement. Thus the chiral operator ${\cal X}$ can serve as a witness
of entanglement~\cite{Horodecki}. An operator ${\hat O}$ is a suitable
entanglement witness if for all separable states $\ket{\Psi}$ its
expectation value $\langle {\hat O} \rangle
\equiv \langle \Psi | {\hat O} | \Psi\rangle$ is bounded and only
entangled states can yield expectation values that exceed this bound.

In our case, the appropriate definition of the \textit{chiral
entanglement witness} is
\begin{equation}
  \label{E_X}
  E_{\cal X} (\Psi) = -1 + \max_{U} |\chi_{i,j,k}|
\end{equation}
where the maximization is over all unitary operators $U\in SU(2)$ acting
locally on each of the spins $i,j,k$. If $E_{\cal X} (\Psi)>0$ then
$\ket{\Psi}$ is guaranteed to be entangled, otherwise we cannot infer the
presence of entanglement on this basis alone. It should be stressed that
$E_{\cal X}(\Psi) > 1$ implies the existence of tripartite entanglement, and
that, in contrast to previous other entanglement witnesses~\cite{Toth}, it
can distinguish between some of the GHZ- and W-entangled states, as shown in
the previous subsection. Finally, note that the definition of $E_{\cal
\chi}$ can be straightforwardly generalized to mixed states.

%%%%%%%%%%%%%%%%%%%%%%%%%%%----------- new sec ------------
\section{Quasi-one dimensional geometries}

\label{sec:ladder}

In this section we consider quasi-one dimensional (1D) geometries
of ladders and polygons as a first step towards studying the two
dimensional (2D) case. We will estimate the ground state
properties of Hamiltonian (\ref{Ham}) using a mean field
variational estimate and verify that it agrees with the outcome of
exact numerical diagonalizations.

\subsection{Different types of quasi-1D spin systems}

\begin{figure}
\centering
\resizebox{0.6\linewidth}{!}{\includegraphics{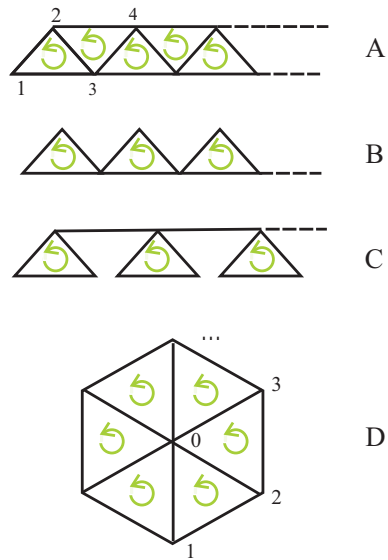}}
\caption{Quasi-1D geometries. The spins reside on the vertices and
spin-spin interactions are represented by the edges. At each
triangle there is also a chiral three-spin interaction which has
the same sense everywhere, as indicated by the arrows.
\label{fig_1}}
\end{figure}

We will particularize the model of Eq.~(\ref{Ham}) to four different
geometries which include three types of spin ladders and a ring-type setup
(see Fig.~\ref{fig_1}). It is useful to reparameterize the Hamiltonian as
follows
\begin{equation}
H = - \sum_{i<j} J_{i,j} \vec\sigma_i \cdot \vec\sigma_j + \lambda
\sum_{i<j<k} X_{i,j,k} \vec\sigma_i \cdot \vec\sigma_j \times
\vec\sigma_k. \label{H}
\end{equation}
Here $J_{i,j}$ and $X_{i,j,k}$ are nonzero only for the appropriate bonds of
each particular lattice and determine both the geometry of the lattice and
the sign of the Heisenberg and chiral terms ($J_{i,j}, X_{i,j,k} \in
\{0,+1,-1\}$).

%%%
%------------------------   F I G   1

For $|\lambda| \ll 1$, the ferromagnetic interaction dominates and
spins align along any direction. For $|\lambda|\gg 1$, the chiral
interaction dominates, the magnetic ordering disappears and we
observe the establishment of chiral order in different plaquettes.
Note that by fixing the sign of $X_{ijk}$ we impose the condition
that all plaquettes have the same chiral orientation [See
Fig.~\ref{fig_1}]. This leads to an additional frustration in the
chiral regime and may prevent the system from achieving the
largest value of the chirality, which is $2\sqrt{3}$ per
plaquette, even for very large values of $\lambda$.

Let us now look at the different quasi-1D models. For the ladder
of type A the nonzero couplings are
\begin{eqnarray}
J_{i,i+1} &=& 1, \\
J_{i,i+2} &=& 1, \nonumber\\
X_{i,i+1,i+2}&=& (-1)^i \nonumber,
\end{eqnarray}
where the factor $(-1)^i$ fixes the same orientation of the chiral
term on all triangles. The ladder of type B is very similar, but
the upper row of Heisenberg and half of the chiral interactions
are missing, so that
\begin{eqnarray}
J_{i,i+1} &=& 1, \\
J_{2i,2i+3} &=& 1, \nonumber \\
X_{2i,2i+1,2i+2}&=& 1. \nonumber
\end{eqnarray}
By contrast, the ladder of type C is formed by a set of weakly
connected triangles, in which case we obtain
\begin{eqnarray}
J_{i,i+1} &=& 1, \quad i=1,2,4,5\ldots\\
J_{i,i+2} &=& 1, \quad i=1,4,7,\ldots\nonumber\\
J_{i,i+3} &=& 1, \quad i=2,5,8,\ldots\nonumber\\
X_{i,i+1,i+2}&=& 1, \quad i=1,4,7,\ldots \nonumber
\end{eqnarray}

Finally, in the ring-type geometry shown in Fig. \ref{fig_1}(D) there is a
central spin-$\frac{1}{2}$ particle (the spin labelled `0') which is
connected to $N-1$ equidistant spins. The Hamiltonian of this system is best
written explicitly as
\begin{eqnarray} \label{H_ring}
H_{\rm ring} &=& - \sum_{i=1}^{N} \vec{\sigma}_0 \cdot
\vec{\sigma}_i - \sum_{i=1}^{N-1} \vec{\sigma}_i \cdot
\vec{\sigma}_{i+1} \nonumber \\
&& + \lambda \sum_{i=1}^{N-1} \vec{\sigma}_0 \cdot \vec{\sigma}_i
\times \vec{\sigma}_{i+1}.
\end{eqnarray}

\subsection{Analytic results}
\label{fermionization}

We have studied analytically these models, in order to get information about
the ground state energies and the location of possible quantum phase
transitions. We present in detail the method only for the type-A ladder, but
the procedure is the same for the rest of the geometries. Our solution works
in three steps (i) a mapping to a hard-core bosonic problem, (ii) a second
mapping to a fermionic model and (iii) a mean-field solution of the
fermionic model.

First of all, the spins are mapped onto bosons using the
Dyson-Maleev transformation \cite{Sachdev},
\begin{eqnarray} \label{Dyson_Maleev}
  X_n = b_n +b^{\dagger}_{n},\;\; Y_n=-i(b_n -b^{\dagger}_{n}), \;\;
  Z_n=1-2b^{\dagger}_{n} b_n \;\;\;
\end{eqnarray}
where the $b, b^{\dagger}$ are bosonic operators that satisfy the
commutation relations $[b_{n}, b^{\dagger}_{m}] = \delta_{nm}$,
and $[b_{n}, b_{m}] = [b^{\dagger}_{n}, b^{\dagger}_{m}] = 0$. To
be consistent with the spin-$\frac{1}{2}$ problem at hand, we
allow up to one boson per lattice site, i.e.,
$b^{\dagger}_{n}b_{n} \le 1$. In this representation the
Heisenberg interaction between the $m$-th and $n$-th spins reads
\begin{eqnarray} \label{Heis_term_ini}
  \vec{\sigma}_m\cdot \vec{\sigma}_n  = (1 - 2b^\dagger_mb_m)
  (1-2b^\dagger_n b_n) + 2(b_m^\dagger b_n + b_n^\dagger b_m)
  \nonumber \\
\end{eqnarray}
while the chiral interaction between spins $l$, $m$ and $n$
becomes
\begin{eqnarray} \label{chiral_term_ini}
  {\cal X}_{lmn} &=& 2i[
  (b_m^\dagger b_l - b_l^\dagger b_m)(1- 2b_n^\dagger b_n)
  \nonumber \\
  && + (b_n^\dagger b_m - b_m^\dagger b_n) (1- 2b_l^\dagger b_l)
  \nonumber \\
  && + (b_l^\dagger b_n - b_n^\dagger b_l) (1- 2b_m^\dagger
  b_m)].
\end{eqnarray}

The next step is to turn to fermionic variables, $c_n$. We achieve
this by using the Jordan-Wigner transformation
\begin{eqnarray}
  \label{Jordan_Wigner}
  c_{n} = (-1)^{\zeta_n} b_n, \quad
  \zeta_n = \sum_{k<n}c_{k}^{\dagger} c_{k}
\end{eqnarray}
where $\zeta_n$ gives the total number of fermions on the sites to
the left of site $n$. These new operators satisfy fermionic
anticommutation relations $\{c_{n}, c^{\dagger}_{m}\} =
\delta_{nm}$, and $\{c_{n}, c_{m}\} = \{c^{\dagger}_{n},
c^{\dagger}_{m}\} = 0$. In addition to this we also need the
following relations
\begin{eqnarray}
  \label{Jordan_Wigner_relations}
  b_{n}^{\dagger} b_{n} = c_{n}^{\dagger} c_{n},\quad
  b_{n}^{\dagger} b_{n+1} = c_{n}^{\dagger} c_{n+1}, \nonumber \\
  b_{n}^{\dagger} b_{n+2} = c_{n}^{\dagger} (-1)^{c_{n+1}^\dagger c_{n+1}}
  c_{n+2}.
\end{eqnarray}
Replacing all formulas above into the particular Hamiltonian
(\ref{H}) one obtains a complicated fermionic model which in
general has no simple analytical solution. We may nevertheless
estimate variationally the properties of the ground state.

Our approximation method proceeds as follows. We notice that for
$|\lambda|\ll 1$ one of the possible ferromagnetic ground states
is polarized along the z-direction. This means that the state has
no effective fermions and $\alpha \equiv \langle b^\dagger_n
b_n\rangle = 0$. It thus makes sense to treat $\alpha$ as our
order parameter; to accurately find the phase transition the
expression as a function of $\alpha$ must become exact in the
limit of one boson, $\alpha = {\cal O}(1/N)$, where the transition
to the chiral regime occurs.

Our mean field method dictates two approximations. In the
Heisenberg term we replace
\begin{equation}
  (1-2c_n^\dagger c_n)(1-2c_m^\dagger c_m) \to
  1 + 2(\alpha-1) (c_n^\dagger c_n + c_m^\dagger c_m)
\end{equation}
for the various pairs of interacting sites. In the chiral term we
simply substitute $(2c_{i+2}^\dagger c_{i+2}- 2 c_{i-1}^\dagger
c_{i-1})$ with 0, which leads to a cancellation of the diagonal
hopping terms, leaving only the quadratic next-nearest-neighbor
hopping. With this, the Hamiltonian of type-A ladder becomes
\begin{eqnarray} \label{H_A_boson}
\lefteqn{ H_{\rm A} = - 2 N
  - 2 \sum_l \left[c_{l+1}^\dagger c_l + c_l^\dagger c_{l+1}
  - 8 (\alpha-1)c_l^\dagger c_l \right]} \nonumber \\
  && - 2\sum_l \left\{[(1-2\alpha)+\lambda i (-1)^l] c_{l+2}^\dagger c_l
  + \mathrm{H.c.} \right\}.
\end{eqnarray}
It is now convenient to regroup the fermionic operators to form a
spinor, $\Psi_l^T\equiv(c_{2l+1},c_l)$, and then to perform a
discrete Fourier transform
\begin{equation}\label{Fourier_transform}
\Psi_l = \frac{1}{\sqrt{N}} \sum_p e^{ip l} \Psi_p
\end{equation}
where $p=2\pi n/L$ for $n=0,1,\ldots (L-1)$ and $L=N/2$ is the
length of the ladder. Consequently the Hamiltonian becomes
\begin{eqnarray}
  H_{\rm A} = - 2N +
  \sum_p \Psi_{p}^{\dagger} M_p \Psi_{p} \label{H_A_diag}
\end{eqnarray}
with the coupling matrix
\begin{eqnarray}
\lefteqn{M_p = [8(1-\alpha)-4(1-2\alpha)\cos(p)] \mathbb{I} } \nonumber \\
  && -2[1+\cos(p)] \sigma_x - 2 \sin(p)\sigma_y - 4\lambda\sin(p)\sigma_z.
\end{eqnarray}
The eigenvalues of the matrix yield the eigenenergies of the
Hamiltonian, which are
\begin{eqnarray} \label{energies_A}
\lefteqn{E_{\pm}(p) = 8(1-\alpha)-4(1-2\alpha)\cos(p)} \nonumber \\
&& \pm 2\sqrt{(4\lambda^2 + 1)\sin^2(p)+ [1+\cos(p)]^2}.
\end{eqnarray}
If we label as $E_k$ the $2L$ eigenenergies, sorted such that $E_k
\leq E_{k+1}$, the lowest energy state of our model has $\alpha N$
fermions occupying the states $k=1$ up to $k=\alpha N$. The energy
is then given by $E=\sum_{k=1}^{\alpha N}$ and the right value of
$\alpha$ is found by solving numerically the self-consistency
equation
\begin{equation}
  \alpha N = \sum_{E_+(p)\leq 0} 1 + \sum_{E_-(p)\leq 0} 1.
\end{equation}

This procedure can be repeated for all the lattice configurations shown in
Fig.~\ref{fig_1}, obtaining again upper bounds for the energy of the ground
state. In addition, we can compute the value of $\lambda$ at which the phase
transition from the ferromagnetic ground state to the chiral state happens.
This is given by the configuration at which $\alpha\neq 0$ becomes
favorable, which is the configuration in which some $E_{\pm}(p)$ become
negative.

\subsection{Numerical results}

%%%
%------------------------   F I G   2
\begin{figure}
\centering \resizebox{\linewidth}{!}{\includegraphics{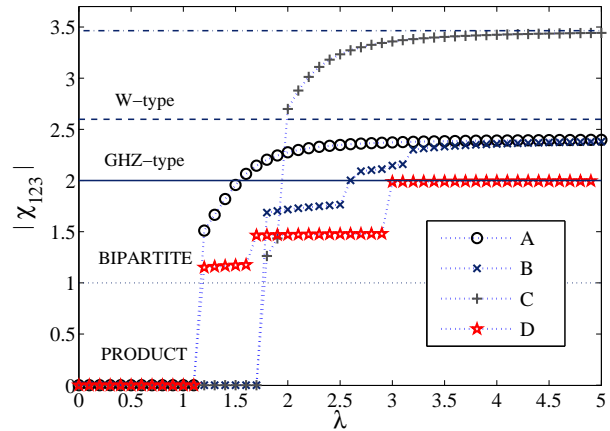}}
\caption{Absolute value of the chirality vs. strength of chiral
interaction for the different geometries of Fig. \ref{fig_1} (for $N=9$
spins with periodic boundary conditions). We see that the chirality of
lattice types A, B, C, and D becomes nonzero at the points $\lambda_{\rm A}
\approx 1.1$, $\lambda_{\rm B} \approx 1.7$, $\lambda_{\rm C} \approx 1.7$,
and $\lambda_{\rm D} \approx 1.1$, respectively. The horizontal lines (from
top to bottom) correspond to the maximum chirality ($=2\sqrt{3}$), and the
entanglement witnesses for W-like ($=3\sqrt{3}/2$), GHZ-like ($=2$) and
bipartite entangled ($=1$) states, as demonstrated in Section II. }
\label{fig_2}
\end{figure}
%%%

%%%
%------------------------   F I G   3
\begin{figure}
\centering \resizebox{\linewidth}{!}{\includegraphics{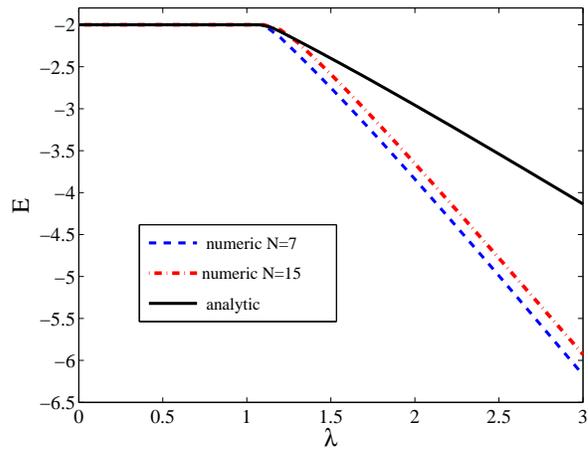}}
\caption{Ground state energy per lattice site vs. strength of
chiral interaction for a type-A ladder (Fig.~\ref{fig_1}a). The solid line
is the analytical estimate computed using the fermionization in
Sect.~\ref{fermionization}, while the dashed and dashed-dotted lines come
from an exact diagonalization of the model for $N=7$ and $N=15$ spins,
respectively.
\label{fig_3}}
\end{figure}
%%%

In addition to the mean field studies, we have diagonalized
exactly the Hamiltonian in Eq.~(\ref{H}) using all four geometries
and an increasing number of spins. In all cases we observe a phase
transition from a ferromagnetic regime ($\lambda=0$) to a chiral
one ($|\lambda|\to\infty$). The location of the phase transition
depends on the particular model and agrees with the estimates
developed above. From the analytics, for the ladder of type A and
for the ring geometry we find $\lambda_{\rm tran} \approx 1.1$,
while for the two other configurations we have $\lambda_{\rm tran}
\approx 1.7$. As shown in Fig.~\ref{fig_2}, these values are close
to the actual location of the jumps in the chirality that are
obtained numerically.

It is interesting to note that, in the chiral phase, the total chirality
strongly depends on the geometry of the lattice. Since our Hamiltonian
favors the same orientation of chirality on all plaquettes, the type-A
ladder experiences a particular kind of frustration, where no two adjacent
plaquettes are able to host maximal chiral currents. A similar phenomenon is
experienced by the type-B lattice, but not by the type-C ladder, where the
weakly connected plaquettes are able to saturate the maximum value of the
chirality per site (\emph{cf}. Fig.~\ref{fig_2}). On the other hand, the
ring configuration appears to have only bipartite entanglement. This is due
to rotational symmetry, which forces the central spin to decouple from each
of the triangles, thus forbidding any possible tripartite correlations.

As we have seen, chirality acts as an entanglement witness for
two- and three-partite entanglement. Following our previous
analysis, and from Fig.~\ref{fig_2}, we conclude that plaquettes
in the A and B ladders have genuine three-partite entanglement of
GHZ-type. Ladder C, with its weakly bound triangles and no
frustration, achieves W-type entanglement on every three
neighboring spins.

We have compared the energies per site provided by the numerical simulation
with the analytical mean-field results. As shown in Fig.~\ref{fig_3}, the
derivative of the energy per site experiences a discontinuity around the
transition point. This feature is shown both in the numerics and in the mean
field estimates. The latter, however, only provide a variational upper bound
that is approached by the numerical exact diagonalization for increasing
number of spins.

\begin{figure}
\centering
\resizebox{0.8\linewidth}{!}{\includegraphics{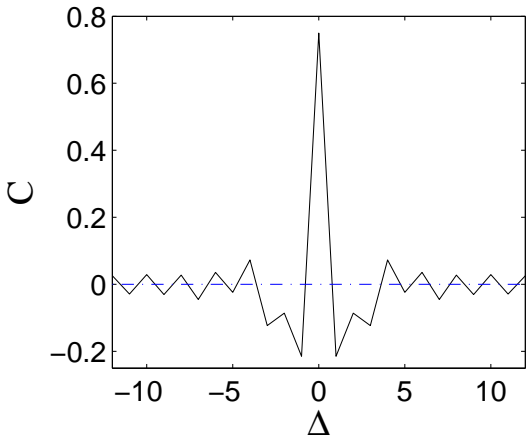}}

\caption{Spin-spin correlations $C_{0,\Delta}$  [Eq.
(\ref{correlator})] for a periodic type-A ladder with $N=24$ spins. The
dashed-dot line corresponds to $\lambda = 0.2$ and the solid line to
$\lambda = 100$. In the chiral regime the correlations decay exponentially
(on either direction of the periodic ladder).
\label{fig_4}}
\end{figure}

We have also investigated numerically the spin-spin correlations
of the ground state of type-A ladder for various ladder sizes and
chiral couplings, $\lambda$, using periodic boundary conditions. A
suitable way to measure the quantum correlations is to evaluate
the connected correlator
\begin{equation}
C_{i,j} = \langle \vec{\sigma}_i \cdot \vec{\sigma}_j \rangle -
\langle\vec\sigma_i\rangle\cdot\langle\vec\sigma_j\rangle
\label{correlator}
\end{equation}
between any two spins.  Scaling studies show that two different behaviors
corresponding to the two phases of the system are obtained. Below the phase
transition, $\lambda < \lambda_{\rm tran} \approx 1.1$, on the ferromagnetic
regime, spins are completely aligned and the connected correlation becomes
zero. Above the phase transition, in the chiral phase $\lambda >
\lambda_{\rm tran} \approx 1.1$ the quantum spin-spin correlations
appear to decrease exponentially fast with the separation of spins, $|i-j|$.
Both regimes are shown in Fig.~\ref{fig_4}, where we plot the correlator
$C_{0,\Delta}$ between the first and any other spin on a type-A ladder with
$N=24$ sites. For $\lambda = 0.2$ the correlator $C_{0,\Delta}$ is
identically zero, while for $\lambda = 100$ it becomes negligible beyond the
sixth site. This demonstrates the absence of structure in the ground state
of type-A ladder when it is in the chiral regime.

%------------------------   F I G   4
%%%

%%%%%%%%%%%%%%%%%%%%%%%%%%%----------- new sec ------------
\section{Two-dimensional model}

\label{sec:torus}

\begin{table}
  \centering
  \begin{tabular}{|c|c|c|}
    \hline
    Rows $\times$ columns & $\ket{S,S_z}$  & degeneracy \\
    \hline \hline
    $2\times n$ & $\ket{0,0}$ & 1 \\
    $3\times 3$ & $\ket{\frac{1}{2},\pm\frac{1}{2}}$ & 4 \\
    $3\times 4$ & $\ket{0,0}$ & 1 \\
    $3\times 5$ & $\ket{\frac{1}{2},\pm\frac{1}{2}}$  & 4 \\
    $4\times 4$ & $\ket{0,0}$ & 1 \\
    $4\times 5$ & $\ket{0,0}$ & 1 \\
    %$5\times 5$ & $\ket{\frac{1}{2},\pm\frac{1}{2}}$ & 4\\
    \hline
  \end{tabular}
  \caption{
    Properties of the ground state in the chiral phase, $|\lambda| \gg 1$,
    for various lattice sizes. We show the magnetization, $S_z$, the total
    spin, $S$, and the degeneracy of the ground state.
  }
  \label{tab:2d}
\end{table}

\begin{figure}[t]
  \centering
  \resizebox{0.8\linewidth}{!}{\includegraphics{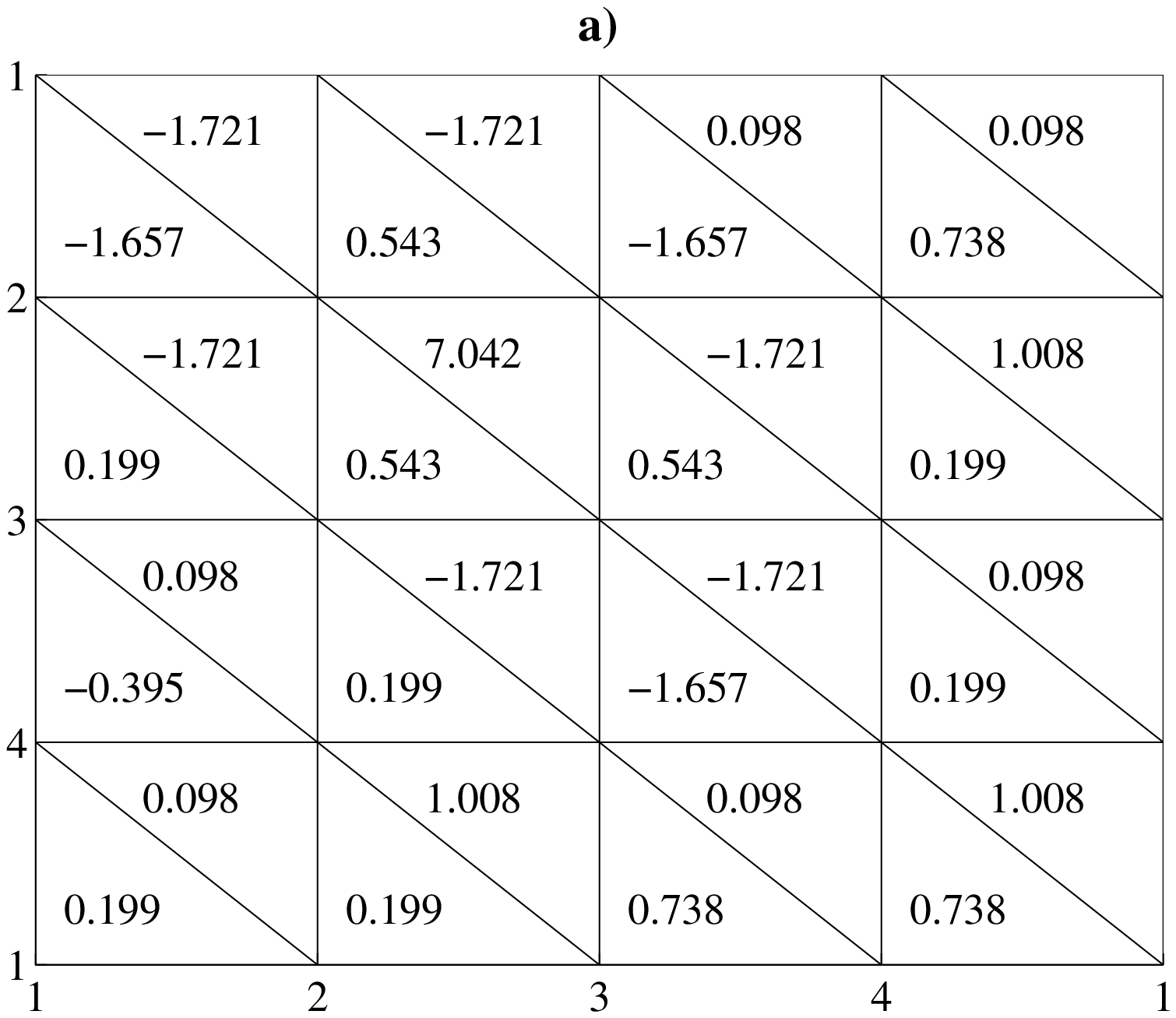}}
  \resizebox{0.8\linewidth}{!}{\includegraphics{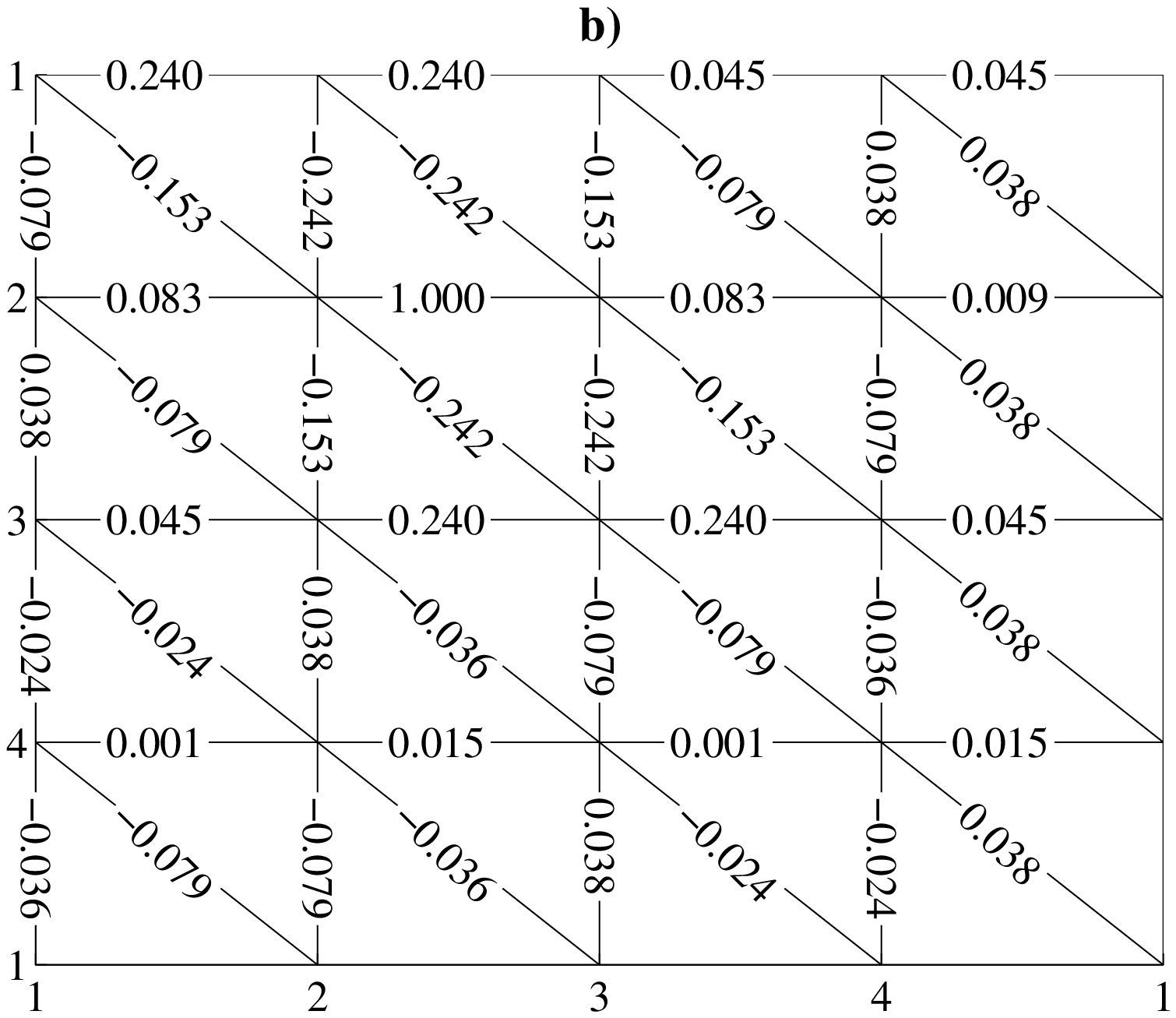}}
  \caption{A geometrical configuration of the $4\times 4$ periodic lattice.
The position $(n,m)$ of the site on the lattice is given in terms of the
integers $n,m=1,...,4$. A certain bond is depicted as $(n,m)-(k,l)$ that
connects the corresponding two neighboring sites. (a) Chiral correlation
between the plaquette with sites $(2,2)$, $(2,3)$ and $(3,3)$ and the
remaining ones. Note the difference between the value of $\langle {\cal XX}
\rangle$ for the same plaquette ($7.042$) and neighboring ones. (b) Dimer
correlations between the bond $(2,2)-(2,3)$ and the remaining bonds. The
simulations were performed deep in the chiral regime $|\lambda|=100$.}
\label{fig_6}
\end{figure}

In this section we consider the two dimensional triangular lattice and
impose periodic boundary conditions in both directions. The lattice is
represented by an array, as illustrated in Fig.~\ref{fig_6}, with dimensions
denoted as N times M. We have performed exact numerical diagonalization of
the Hamiltonian for various sizes of the lattice and characterized both the
ground state and its lowest excitations.

\begin{figure}
  \centering
  \resizebox{0.8\linewidth}{!}{\includegraphics{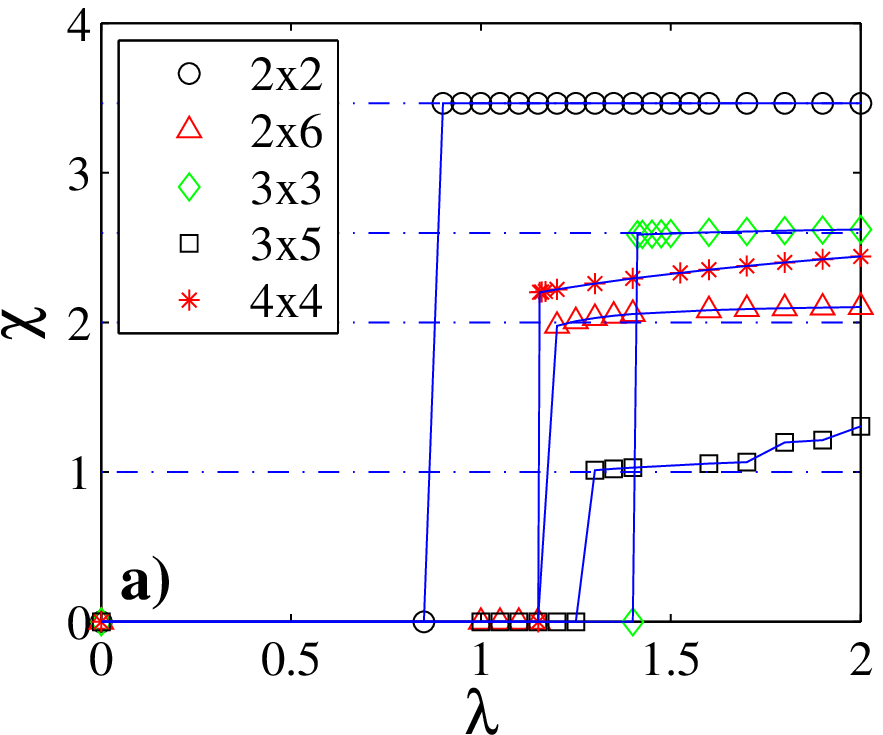}}
  \resizebox{0.8\linewidth}{!}{\includegraphics{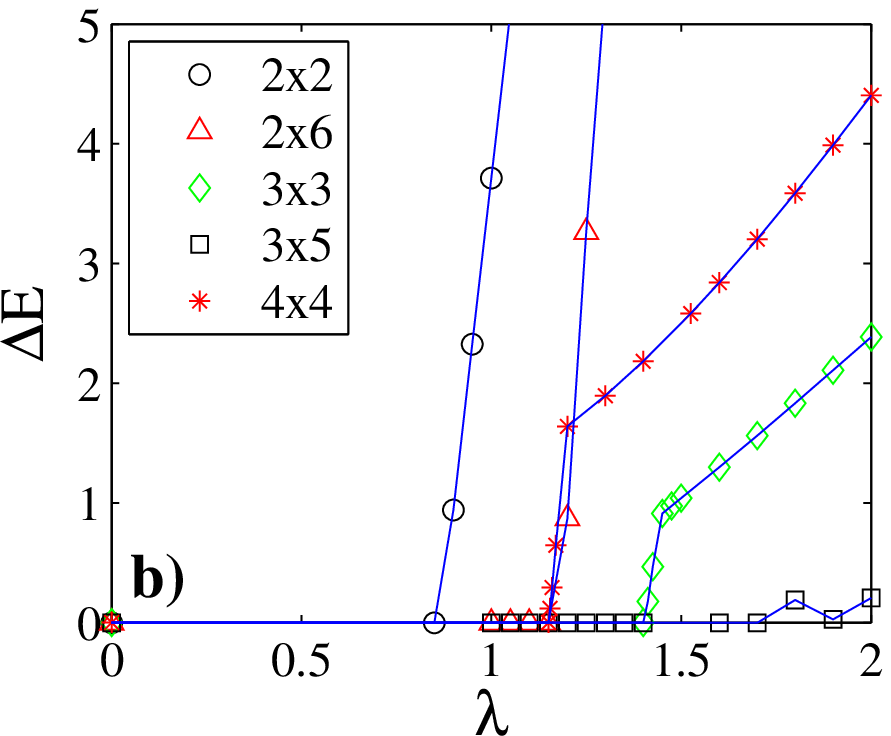}}
  \caption{(a) Ground state mean chirality as a function of
the coupling constant $\lambda$ for different lattice sizes. (b) Energy gap
between the ground state and the first excited state.}
\label{fig_5}
\end{figure}

The results are summarized in Table~\ref{tab:2d} and in Fig.~\ref{fig_5}.
Once more, we obtain two phases separated by a phase transition point at a
position that approaches $\lambda_{\rm tran} \approx 1.1$ for large
lattices. When the chiral coupling $\lambda$ is small ($\lambda<\lambda_{\rm
tran}$) then the Heisenberg interaction is dominant and the system is in a
ferromagnetic state. When the coupling is large ($\lambda>\lambda_{\rm
tran}$) then the chiral term is dominant causing the ground state to have
non-zero chirality (see Fig.~\ref{fig_5}(a). Connecting with our studies of
the chirality as an entanglement witness, the larger degree of frustration
in the toric setup prevents the system from reaching a large value of the
chirality, which lays around $1.5$ in the limit $|\lambda|\to\infty$. This
implies that the state belongs to the set of states which are at least
two-party entangled.

%%%
%------------------------   F I G   5
%%%

The existence of the quantum phase transition and even its location agree
qualitatively with the features of the Type A ladder, whose geometry most
closely resembles that of these toric structures. However, as we show below,
there are some differences. The most notable one is in the angular momentum
of the ground state. We have found that in the chiral phase the system tends
to adopt the state with the lowest total angular momentum which is
compatible with the number of spins. Thus, as shown in Table~\ref{tab:2d},
if the number of spins is even, which includes the case of Type A ladders,
the ground state is a state with $S=0$ and $S_z=0$ and has no degeneracy.
The corresponding momentum of the ground state is zero. However, if the
total number of spins is odd, the total angular momentum must be fractional,
having $S=\frac{1}{2}$ and $S_z=\pm\frac{1}{2}$. In this case the ground
state becomes four-fold degenerate distinguished by the $z$ component of the
spin and by the momentum which is either zero or $\pi$ in both directions.

It is interesting to note that the excited states almost always have a
larger angular momentum and are separated by a large energy gap, ${\cal
O}(|\lambda|)$, from the ground state. This is illustrated in
Fig.~\ref{fig_5}(b). This energy gap, which survives in the thermodynamic
limit, is responsible for a finite correlation length. This causes both the
spin-spin and chiral-chiral correlations to die off exponentially fast (see
Fig.~\ref{fig_4} for the spin-spin correlations in the case of a type-A
ladder).

Remarkably, on some special cases, such as the $3\times 5$ lattice, the gap
between the ground and excited states is much smaller (see
Fig.~\ref{fig_5}(b)). The excited states then have similar values of total
angular momenta and their chiralities differ only slightly --- by $0.2$ or
less --- from that of the ground state. We conjecture that the existence of
these states is supported by the frustration of the chirality in
Eq.~(\ref{Ham}) where we enforce all plaquettes to have a similar
orientation of the spin current.

The chiral-chiral correlations are presented in Fig.~\ref{fig_6} for a
$4\times4$ lattice, where we show the connected correlation
\begin{equation}
  \langle {\cal XX} \rangle \equiv
  \langle {\cal X}_{ijk} {\cal X}_{lmn}\rangle
  - \langle {\cal X}_{ijk}\rangle\langle {\cal X}_{lmn}\rangle
\end{equation}
between a reference plaquette $(lmn)$ and any other one, $(ijk)$. In
Fig.~\ref{fig_6}(a) the reference plaquette is given by the sites $(2,2)$,
$(2,3)$ and $(3,3)$ of the $4\times 4$ lattice. The value of $\langle {\cal
XX} \rangle$ between the reference site and itself is $7.042$, while the
connected correlation with respect to neighboring triangles decreases
dramatically. This indicates the absence of any structure in the chirality
of the ground state.

%%%
%------------------------   F I G   6
%%%

The structure of the ground state is a singlet whenever the total number of
spins is even. This by itself does not necessarily imply a complicated
hidden order, since we could achieve the same value of the total angular
momentum by packing the spins into singlets. Following
Ref.~\cite{misguich98} we have computed the dimer order parameter, which is
defined as the connected correlation of an operator $d_{ij} = (1
-\vec\sigma_i \cdot\vec\sigma_j)/4$ that detects singlets. The dimer-dimer
correlations are then defined as
\begin{equation}
  D_{ij} = \frac{\langle d_{ij}d_{kl}\rangle - \langle d_{ij} \rangle
    \langle d_{kl}\rangle}{\langle d_{kl}\rangle (1 -\langle d_{ij}\rangle)},
\end{equation}
where $(kl)$ is the reference bond and $(ij)$ any other bond.  In
our case the value of the dimer order $D$ decreases rapidly as a
function of the bond separation (see Fig.~\ref{fig_6}a),
suggesting that this is not a dimer solid.

Summarizing, the ground state of the $4\times 4$ periodic lattice appears to
have no structure with respect to spin-spin correlations or
chirality-chirality correlations. Moreover, the dimer correlations appears
to reduce dramatically for increasing distance between the dimers.

\section{Optical lattice implementation}

\label{sec:implementation}

\begin{figure}
\centering \resizebox{!}{3.2cm}{\includegraphics{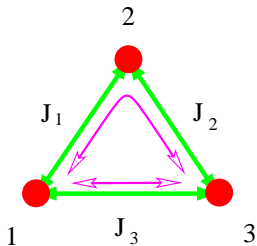}}
\caption{\label{triangle} The triangular configuration. Three-spin
    interaction terms appear between sites $1$, $2$ and $3$ since
    tunnelling between two of the sites, such as $1$ and $2$,
    can happen directly or through the third site.
    The latter case results into an exchange interaction between
    $1$ and $2$ that is influenced by the state of the spin at site $3$.}
\end{figure}

Here we would like to sketch a method of how to produce the Heisenberg and
chiral interactions studied in this paper by employing cold atoms superposed
with optical lattices. More precisely we consider a Mott insulator of two
species bosonic atoms loaded on an optical lattice with a triangular
configuration, as shown in Fig.~\ref{triangle}. These two states can be
encoded in the internal hyperfine states of an atom. In the limit of deep
off-resonance optical lattices, the evolution of this system is described by
a Bose-Hubbard Hamiltonian for atoms of species ${\cal S}=a,b$
\begin{eqnarray}
H &=& -\sum_{i,{\cal S}} (J^{{\cal S}}_{i} a_{i,{\cal S}}^\dagger
a_{i+1,{\cal S}}
+ {\rm H.c.}) \nonumber \\
&+& \frac{1}{2} \sum _{i, {\cal S}} U_{{\cal S}, {\cal S}} n_{i,{\cal
S}}(n_{i,{\cal S}}-1) + \sum _{i} U_{a,b} n_{i,a}n_{i,b}
\end{eqnarray}
Here $J^S>0$ is the atom tunnelling coupling between neighboring sites,
$U_{S,S'}$ is the on-site interaction between atoms $S$ and $S'$,
$a_{i,{\cal S}}$ denotes the annihilation operator of atoms of species
${\cal S}$ at the site $i$ and $n_{i,{\cal S}}$ is the corresponding number
operator. For simplicity we take $U_{S,S'}$ to be positive.

%------------------------   F I G   7

As we are interested in a chiral interaction we want to generate an
effective charge-magnetic field coupling in the atomic system manifested by
complex, position dependent, tunnelling couplings. For example, it has been
shown in Refs.~\cite{Pachos,Kay,Pachos_Rico} that a gradient of a magnetic
field along the direction of the electric dipole moment can simulate
effectively the interaction between a charge and a homogenous magnetic
field. Alternatively, Raman assisted tunnelling with lasers that have a
phase difference~\cite{Jaksch} can be also employed. We will employ such a
technique here to generate, in a controlled way, complex position dependent
tunnelling couplings.

In the insulator regime, hopping is weak compared to the interaction, $J \ll
U$, and atoms have a low probability of jumping to other sites. For
populations of only one atom per lattice site we can employ the pseudo-spin
basis for each lattice site, given by $|n_{a}=1,n_{b}=0\rangle\equiv
|\!\!\uparrow\rangle $ and $|n_{a}=0,n_{b}=1\rangle\equiv
|\!\!\downarrow\rangle $. In that limit one can treat the hopping terms
perturbatively and find that they help to establish effective exchange
interactions between atoms in neighboring sites. Up to third order in the
perturbation expansion the effective Hamiltonian is given by
\begin{eqnarray} \label{complexboson}
H &=& \alpha\sum_i I\!\!I +\beta\sum Z_i +\gamma \sum_i
Z_i Z_{i+1} \nonumber \\
&+& \delta \sum_i (X_i Y_{i+1}- Y_i X_{i+1}) \nonumber \\
&+& \epsilon
\vec{\sigma}_i\cdot\vec{\sigma}_{i+1}\times\vec{\sigma}_{i+2}
\end{eqnarray}
The presented couplings are given by
\begin{eqnarray}
&&\lefteqn{ \alpha = \frac{{J^a}^2 } { U_{aa}} + \frac{{J^b}^2} {U_{bb}},
\,\,\,\,\, \beta = \frac{{J^a}^2}{2U_{aa}} - \frac{{J^b}^2 }{2 U_{bb}} }
\nonumber \\
&& \gamma= \frac{{J^a}^2}{U_{aa}} + \frac{{J^b}^2}{U_{bb}},
\,\,\,\,\, \delta= i\left(\frac{{J^a}^2
J^b}{2U_{aa}^2} + \frac{{J^b}^2 J^a}{2U_{bb}^2}\right) \nonumber \\
&& \epsilon= i\left(\frac{{J^a}^2 J^b}{2U_{aa}^2} - \frac{{J^b}^2
J^a}{2U_{bb}^2}\right). \nonumber
\end{eqnarray}
If we choose $J^b$ to be a negative imaginary number then it is possible to
set $\alpha$, $\gamma$ and $\delta$ equal to zero. In addition, we can apply
an effective magnetic field that cancels the $\beta$ term, resulting
eventually to an isolated $\epsilon$ term,
\be
H=\epsilon\sum_i\vec{\sigma}_i\cdot(\vec{\sigma}_{i+1} \times
\vec{\sigma}_{i+2}).
\ee
This is the chiral interaction that we required.

The presented interaction is produced from third order perturbation theory
and it could be rather small~\cite{PachosKnight}. A similar interaction can
be engineered using recently developed techniques that involve cold
molecules in optical lattices~\cite{Zoller}. Compared to the cold atom
implementation shown here, the molecules would have the advantage of
producing stronger interactions.

\section{Conclusions}

\label{sec:summary}

In this article we have studied a chiral spin system which is
defined on a two dimensional triangular lattice. The analytical or
numerical study of the 2D-model is hard so we initially resorted
to simplified geometries such as triangular ladders and polygons
(Fig.~\ref{fig_1}). By fermionizing and using mean field
approximation we obtained the transition points, that separate
between a spin-ordered and a chiral phase, as well as an upper
bound for the ground state energies. These are in agreement with
exact diagonalization of finite size systems (see
Figs.~\ref{fig_2},~\ref{fig_3}). From these results we have
observed that for the ladder of type A the phase transition
happens for smaller values of the chiral coupling $\lambda$,
compared to the corresponding transition points for cases B and C.
This has been attributed to frustration effects that also arise in
the two dimensional case. On the contrary, case C has chirality
that saturates the upper bound $\chi\leq2\sqrt{3}$, for large
chiral coupling $\lambda$. Moreover, case A and the 2D-lattice
saturate to a lower value of the chirality. Using the chirality as
a witness of entanglement, we have discussed the entanglement
properties of the system and distinguished between two-party and
genuine three-party entanglement. We have suggested that the
recent advances in cold atom and polar molecule technologies
\cite{Pachos,Kay,Pachos_Rico,Jaksch,Zoller} could facilitate the
experimental realization of the chiral systems presented here.

The study of topological order is rather complex and its presence
in actual physical systems is hard to probe in a definitive way.
Moreover, finite size simulations make it possible to deduce
certain characteristics which can only indicate the presence of
topological order. Indeed, the extensive numerical calculations
carried out here on the 2D model suggest the presence of
topological order in the ground state of the system, which has
been supported by a number of observations. Firstly, the system
exhibits a ground state degeneracy that goes beyond the breaking
of the PT-symmetry. Secondly, there is an energy gap between the
ground and the excited states, which persists with increasing
lattice size. Finally, in the chiral regime the two-point spin
correlations, the chiral connected correlation and the dimer
correlations appear to decay exponentially, a behavior that
indicates the absence of any local structure in the ground states.

%%%%%%%%%%%%%%%%%%%%%%%%%%%----------- thanks! ------------
\acknowledgements DIT would like to thank Susana Huelga, Ray
Bishop, and Hans Mooij for useful discussions. This work was
supported by the Royal Society and the EPSRC (through EP/D065305/1
and the QIP-IRC). J.J.G.-R. acknowledges financial support from
the Ramon y Cajal Program of the Spanish M.E.C. and from the
projects FIS2006-04885 (M.E.C.) and CAM-UCM/910758.

%%%%%%%%%%%%%%%%%%%%%%%%%%%----------- new sec ------------

%----------------------------------------------------------
\end{document}